\documentclass[runningheads]{llncs}

\usepackage{graphicx}

\usepackage{color}
\usepackage{hyperref}

\usepackage{amsfonts}
\usepackage{enumerate}
\usepackage{textcomp}

\begin{document}

\title{Comparing machine learning models \\ to choose the variable ordering for \\ cylindrical algebraic decomposition}
\titlerunning{Comparing machine learning models to choose the variable ordering for CAD}

\author{Matthew England \and Dorian Florescu}
\authorrunning{M. England and D. Florescu}

\institute{Faculty of Engineering, Environment and Computing, \\Coventry University, Coventry CV1 5FB, UK
\email{\\ \{Matthew.England, Dorian.Florescu\}@coventry.ac.uk}}

\maketitle              

\begin{abstract}

There has been recent interest in the use of machine learning (ML) approaches within mathematical software to make choices that impact on the computing performance without affecting the mathematical correctness of the result.  We address the problem of selecting the variable ordering for cylindrical algebraic decomposition (CAD), an important algorithm in Symbolic Computation.
Prior work to apply ML on this problem implemented a Support Vector Machine (SVM) to select between three existing human-made heuristics, which did better than anyone heuristic alone. 
Here we extend this result by training ML models to select the variable ordering directly, and by trying out a wider variety of ML techniques.

We experimented with the NLSAT dataset and the Regular Chains Library CAD function for Maple 2018.  For each problem, the variable ordering leading to the shortest computing time was selected as the target class for ML. Features were generated from the polynomial input and used to train the following ML models: k-nearest neighbours (KNN) classifier,  multi-layer perceptron (MLP), decision tree (DT) and SVM, as implemented in the Python scikit-learn package.  We also compared these with the two leading human-made heuristics for the problem: the Brown heuristic and sotd.  On this dataset all of the ML approaches outperformed the human-made heuristics, some by a large margin. 

\keywords{computer algebra; symbolic computation; non-linear real arithmetic; cylindrical algebraic decomposition; machine learning.}
\end{abstract}

\section{Introduction}

A logical statement is \emph{quantified} if it involves the universal quantifier $\forall$ or the existential quantifier $\exists$.  The \emph{Quantifier Elimination} (QE) problem is to derive from a quantified formula an equivalent un-quantified one.  A simple example would be that the quantified statement, ``$\exists x.\, x^2 + b x + c = 0$'' is equivalent to the unquantified statement ``$b^2 - 4c \geq 0$'', when working over the real numbers.  QE is one definition for simplifying or solving a problem.  The tools involved fall within the field of Symbolic Computation, implemented in Computer Algebra Systems (or more recently in SMT-solvers).  

Our work is on Quantifier Elimination over the reals.  Here the logical statements are expressed as \emph{Tarski formulae}, Bool\-ean combinations ($\land, \lor, \neg, \rightarrow$) of statements about the signs of polynomials with integer coefficients.  QE in this theory was first shown to be soluble by Tarski \cite{Tarski1948} in the 1940s.  However, the only implemented general real QE procedure has algorithmic complexity doubly exponential in the number of variables \cite{DH88}, a theoretical result experienced in practice.  For many problem classes QE procedures will work well at first, but as the problem size increases the doubly exponential wall is inevitably hit.  It is hence of critical importance to optimise QE procedures and the formulation of problems, to ``push the doubly exponential wall back'' and open up a wider range of tractable applications.

QE procedures can be run in multiple ways to solve a given problem: they can be initialized with different options (e.g. variable ordering \cite{DSS04}, equational constraint designation \cite{BDEW13}); tasks can be completed in different orders (e.g. order of constraint analysis \cite{EBCDMW14}); and the problem itself may be expressible in different formalisations \cite{WDEB13}.  Changing these settings can have a substantial effect on the computational costs (both time and memory) but does not effect the mathematical correctness of the output.  They are thus suitable candidates for machine learning: tools that allow computers to make decisions that are not explicitly programmed, via the statistical analysis of large quantities of data.  


We continue in Section \ref{SEC:Background} by introducing background material on the particular decision we study:  the variable ordering for Cylindrical Algebraic Decomposition.  Here we also outline prior attempts to solve this problem; and prior applications of machine learning to computer algebra.
Then in Section \ref{SEC:Methodology} we describe our methodology covering datasets, software, features extracted from the problems, machine learning models tests, and how we test against human-made heuristics.  We present our results in Section \ref{SEC:Results} and final thoughts in Section \ref{SEC:End}.  

\section{Variable Ordering for CAD}
\label{SEC:Background}

\subsection{Cylindrical algebraic decomposition}

A \emph{Cylindrical Algebraic Decomposition} (CAD) is a \emph{decomposition} of ordered $\mathbb{R}^n$ space into cells arranged \emph{cylindrically}: meaning the projections of any pair of cells with respect to the variable ordering are either equal or disjoint.  The cells are (semi)-algebraic meaning each cell can be described with a finite sequence of polynomial constraints.  
A CAD is produced to be invariant relative to an input, i.e. \emph{truth-invariant} for a logical formula (so the formula is either true or false throughout each cell).  Such a decomposition can then be used to perform quantifier elimination on the formula by testing a finite set of sample points and constructing a quantifier-free formula from the semi-algebraic cell descriptions.

CAD was introduced by Collins in 1975 \cite{Collins1975} and works relative to a set of polynomials.  Collins' CAD produces a decomposition so that each polynomial has constant sign on each cell (thus truth invariant for any formula built with those polynomials).  The algorithm first projects the polynomials into smaller and smaller dimensions; and then uses these to lift $-$ to incrementally build decompositions of larger and larger spaces according to the polynomials at that level.  For full details on the original CAD algorithm see \cite{ACM84I}. 

QE has numerous applications throughout science and engineering (see for example the survey \cite{Sturm2006}).  New applications are found regularly, such as the derivation of optimal numerical schemes \cite{EH16}, and the validatation of economic hypotheses \cite{MDE18}, \cite{MBDET18}.  CAD has also found application independent of QE, such as reasoning with multi-valued functions \cite{DBEW12} (where we decompose to see where simplification rules are valid); and biological networks \cite{BDEEGGHKRSW17}, \cite{EEGRSW17} where we decompose to identify regions in parameter space where multi-stationarity can occur. 

The definition of cylindricity and both stages of the algorithm are relative to an ordering of the variables.  
For example, given polynomials in variables ordered as $x_n \succ x_{n-1} \succ \dots, \succ x_2 \succ x_1$ we first project away $x_n$ and so on until we are left with polynomials univariate in $x_1$.  
We then start lifting by decomposing the $x_1-$axis, and then the $(x_1, x_2)-$plane and so so on.  The cylindricity condition refers to projections of cells in $\mathbb{R}^n$ onto a space $(x_1, \dots, x_m$) where $m<n$. 

There have been numerous advances to CAD since its inception, for example: on how best to implement the projection \cite{Hong1990}, \cite{McCallum1998}, \cite{Brown2001a}, \cite{MPP19}; avoiding the need for full CAD \cite{CH91}, \cite{WBDE14}; symbolic-numeric lifting schemes \cite{Strzebonski2006}, \cite{IYAY09}; adapting to the Boolean structure in the input \cite{BDEMW13}, \cite{BDEMW16}, \cite{EBD15};  and local projection approaches \cite{BK15}, \cite{Strzebonski2016}.  However, in all cases, the need for a fixed variable ordering remains.

\subsection{Effect of the variable ordering} 

Depending on the application requirements the variable ordering may be determined, constrained, or entirely free.  The most common application, QE, requires that the variables be eliminated in the order in which they are quantified in the formula but makes no requirement on the free variables.  For example, we could eliminate the quantifier in  
$\exists x.\, ax^2 + b x + c = 0$
using any CAD which eliminates $x$ first; giving six possible orderings to choose from.  A CAD for the polynomial under ordering $a \prec b \prec c$ has only 27 cells, but needs 115 for the reverse ordering.

Note that since we can switch the order of quantified variables in a statement when the quantifier is the same, we also have some choice on the ordering of quantified variables.  For example, a QE problem of the form $\exists x \exists y \forall a \, \phi(x, y, a)$ could be solved by a CAD under either ordering $x \succ y \succ a$ or ordering $y \succ x \succ a$.

The choice of variable ordering has been long known to have a great effect on the time and memory use of CAD, and the number of cells in the output.  In fact, Brown and Davenport presented a class of problems in which one variable ordering gave output of double exponential complexity in the number of variables and another output of a constant size \cite{BD07}.  

\subsection{Prior work on choosing the variable ordering}

Heuristics have been developed to choose a variable ordering, with Dolzmann et al. \cite{DSS04} giving the best known study.  After analysing a variety of metrics they proposed a polynomial degree based heuristic (the heuristic sotd defined later).  However the second author demonstrated examples for which that heuristic could be misled in CICM 2013 \cite{BDEW13}; showed that tailoring it to an implementation could improve its performance in ICMS 2014\cite{EBDW14}; and in CICM 2014 \cite{HEWDPB14} reported that a computationally cheaper heuristic by Brown actually outperforms sotd.

In CICM 2014 \cite{HEWDPB14} we used a support vector machine (SVM), an ML model widely used for non-linear classification, to choose which of three human-made heuristics to believe when picking the variable ordering.  The experiments in \cite{HEWDPB14} identified substantial subclasses on which each of the three heuristics made the best decision, and demonstrated that the machine learned choice did significantly better than any one heuristic overall.  This motivated the present study where we consider a wider range of machine learning models and have these pick the ordering directly from the full range of choices.

\subsection{Other applications of ML to mathematical software}

The CICM 2014 paper was the first to document the application of machine learning to CAD, or in fact to any symbolic computation algorithm / computer algebra system.  Since then there have been two further studies:
\begin{itemize}
\item The same authors studied a different choice related to CAD (whether to precondition the input with Gr\"{o}bner bases) in \cite{HEDP16} \cite{HEWBDP19}, again finding that a support vector machine could make the choice more accurately than the human-made heuristic (if features of the Gr\"{o}bner Basis could be used).
\item At MACIS 2016 there was a study applying a support vector machine to decide the order of sub-formulae solving for a QE procedure \cite{KIMA16}.
\end{itemize}
The survey paper \cite{England2018} and the ICMS Special Session on Machine Learning for Mathematical Software demonstrated the wide range of other potential applications.  As discussed there, while the use of machine learning in computer algebra is rare it has become a key tool in other mathematical software development.  Most notably automated reasoning \cite{Urban2007}, \cite{KBKU13}, \cite{BHP14}, \cite{ACEISU16}; 
but also satisfiability checking \cite{XHHL08}, \cite{LHPCG17}.

\section{Methodology}
\label{SEC:Methodology}

\subsection{Dataset}

Despite its long history and significant software contributions the Computer Algebra community had a lack of substantial datasets \cite{HL15}: a significant barrier to machine learning.  Despite efforts to address this\footnote{E.g. the PoSSo and FRISCO projects in the 90s and the SymbolicData Project \cite{GNJ14}.}, the most substantial dataset of problems designed for CAD is  \cite{WBD12_EX} with less than 100 examples.

However, CAD has recently found prominence in a new area: Satisfiability Modulo Theories (SMT).  Here, efficient algorithms for the Boolean SAT problem study the Boolean skeleton of a problem, with a theory solver then checking if a satisfying Boolean assignment is valid in the domain (learning new clauses if not) \cite{BSST09}.  For the SMT domain of Non-linear Real Arithmetic (NRA), CAD can play the role of such theory solvers\footnote{However, as discussed by \cite{AAB+16a} a more custom approach is beneficial.}, and so their test problems may be used to evaluate CAD.  
We use the \texttt{nlsat} dataset\footnote{\url{http://cs.nyu.edu/~dejan/nonlinear/}} produced to evaluate the work in \cite{JdM12}.  The main sources of the examples are: \textsc{MetiTarski} \cite{Paulson2012}, an automatic theorem prover for theorems with real-valued special functions (it applies real polynomial bounds and then using QE tools like CAD); problems originating from attempts to prove termination of term-rewrite systems; verification conditions from Keymaera \cite{PQR09}; and parametrized generalizations of geometric problems. 

The problems in the \texttt{nlsat} dataset are all fully existential (the only quantifier is $\exists$) which is why they may be studied by SAT solvers.  Although CAD can make adaptions based on the quantifiers in the input (most notably via Partial CAD \cite{CH91}) the conclusions drawn are likely to be applicable outside of the SAT context.  

We extracted $6117$ problems with $3$ variables from this database, meaning each has a choice of six different variable orderings.  We randomly divided them into two datasets for training ($ 4612 $) and testing ($ 1505 $). The training dataset was used to tune the parameters of the ML models. The testing dataset was unknown to the models during training, and is used to compare the performance of the different ML models and the human-made heuristics.

\subsection{Software}

We experimented using the CAD routine \texttt{	
CylindricalAlgebraicDecompose} which is part of the \texttt{RegularChains} Library for \textsc{Maple}.   This algorithm builds decompositions first of $n$-dimensional complex space before refining to a CAD of $\mathbb{R}^n$ \cite{CMXY09}, \cite{CM14b}, \cite{BCDEMW14}.  We ran the code in Maple $ 2018 $ but used an updated version of the \texttt{RegularChains} Library downloaded from \url{http://www.regularchains.org}, which contains bug fixes and additional functionality.  We ignored the quantifiers and logical connectives, using only the polynomials as input to CAD.  The function thus returned a sign-invariant CAD for the polynomials.  

The training and evaluation of the machine learning models was done using the  \texttt{scikit-learn} package \cite{SciKitLearn2011} v0.20.2 for Python 2.7.  The features for machine learning were extracted using code written in the \texttt{sympy} package 
v1.3 for Python 2.7, as was the Brown heuristic.  The sotd heuristic was implemented in \textsc{Maple} as part of the \texttt{ProjectionCAD} package \cite{EWBD14}.

\subsection{Timings}

CAD construction was timed in a Maple script that was
called using the {\it os} package in Python for each CAD, to avoid Maple caching of results.

The target variable ordering for ML was defined as the one that minimises the computing time for a given SAT problem. All CAD function calls included a time limit. The problems in the training dataset were processed with an initial time limit of $4$ seconds on all variable orderings. The time limit was doubled if all $6$ orderings timed out.  The process stopped when the CAD routine was completed for at least one ordering. All problems in the training dataset could be assigned a target variable ordering using time limits smaller than $64$ seconds.

The problems in the testing dataset were processed with a larger time limit of $ 128 $ seconds for all orderings. This was in order to allow a better comparison of the computing times for the ML and the heuristics. When a variable ordering timed out, the computing time was considered equal to $ 128 $ seconds. 

\subsection{Features}

We computed the same set of $ 11 $ features for each SAT problem as \cite{HEWDPB14}, which are listed in Table \ref{tab1}.  
All these features could be computed from the input polynomials immediately.  A possibility for future work is to consider features that are more expensive, such as those from post-processing as in \cite{HEDP16}, those from the end of CAD projection as sotd does, or perhaps even going further into partial lifting as in \cite{WEBD14}.
The ML models associate a predicted variable ordering to each set of $ 11 $ features.  The training and testing feature set were normalised using the mean and standard deviation of the training set.

\begin{table}[hb]
	\caption{The features used by ML to predict variable orderings.}\label{tab1}
	\centering
	\begin{tabular}{|c|cl|}
		\hline
		\textbf{Feature number} & \quad &  \textbf{Description}\\
		\hline
		1 & &  Number of polynomials.\\
		2 & &  Maximum total degree of polynomials.\\
		3 & &  Maximum degree of $ x_0 $ among all polynomials. \\
		4 & &  Maximum degree of $ x_1 $ among all polynomials. \\
		5 & &  Maximum degree of $ x_2 $ among all polynomials.\\
		6 & &  Proportion of $ x_0 $ occuring in polynomials.\\
		7 & &  Proportion of $ x_1 $ occuring in polynomials.\\
		8 & &  Proportion of $ x_2 $ occuring in polynomials.\\
		9 & & Proportion of $ x_0 $ occuring in monomials.\\
		10 & & Proportion of $ x_1 $ occuring in monomials.\\
		11 & & Proportion of $ x_2 $ occuring in monomials.\\
		\hline
	\end{tabular}
\end{table}

\subsection{ML models}

Four of the most commonly used deterministic ML models were tuned on the training data (for details on the methods see for example the textbook \cite{Bishop2006}):
\begin{itemize}
\item The K$-$Nearest Neighbours (KNN) classifier \cite[\S 2.5]{Bishop2006}.
\item The Multi-Layer Perceptron (MLP) classifier \cite[\S 2.5]{Bishop2006}.
\item The Decision Tree (DT) classifier \cite[\S 14.4]{Bishop2006}.
\item The Support Vector Machine (SVM) classifier with RBF kernel \cite[\S 6.3]{Bishop2006}. 
\end{itemize}

The KNN classifier is a type of {\it instance-based} classifier, i.e. it does not construct an internal model of the data but stores all the instances in the training data for prediction. So for each new data instance the model selects the nearest $k$ training instances. Selection can be performed by weighting instances equally, or by weighting a training instance inversely proportional to its distance from the new instance. The prediction is given by the class with the highest count among those training instances. Three algorithms are typically used to train KNN: the {\it brute force} algorithm computes the distances between all pairs of points, the {\it k-dimensional tree} algorithm partitions the data along Cartesian axes, and the {\it ball tree} algorithm partitions the data in a series of nesting hyper-spheres.

The DT is a non-parametric model that uses a tree-like model of decisions and their possible consequences. Each node in the tree represents a test on an attribute, each branch the outcome of the test. The {\it leaves} are the end points of each branch, representing the predicted class label. 
There are  two common criteria used to assess the quality of a split in the DT. The Gini impurity criterion verifies how often a randomly chosen element would be correctly labelled if it were randomly labeled according to the distribution of labels in the subset. The entropy criterion assesses the information gain after each split. 

The SVM is a model that can perform linear and non-linear classification, by selecting an appropriate kernel. It is also known as a maximum-margin classifier, because it identifies a hyperplane in the feature space that maximises the distance to the nearest data points5. The SVM kernel acts as a similarity function between the training examples and a few predefined {\it landmarks}, that can offer additional computing performance. The most common kernels are: {\it linear}, {\it polynomial}, {\it sigmoidal} and {\it radial basis function} (RBF). The RBF kernel is one of the most common kernel choices, given by
\[
\mathcal{K}({\bf x},{\bf \ell})=e^{-\gamma\cdot\|{\bf x}-{\bf \ell}\|^2},
\nonumber
\]
where $ \gamma $ is the kernel hyperparameter, and $ {\bf \ell}\in\mathbb{R}^{n} $ is a predefined {\it landmark}, and $ {\bf x}\in \mathbb{R}^{n} $ is the training vector with $ n $ features.

The MLP is a class of feedforward artificial neural networks. It consists of a minimum of $3$ layers: the input, hidden and output layer. Both the hidden and output layers use a nonlinear activation function that can be selected during cross-validation. Some of the common activation functions are: {\it identity}, {\it logistic}, {\it hyperbolic tangent}, and {\it rectified linear}.

\subsection{Training}

Each model was trained using grid search $ 5 $-fold cross-validation. Specifically, the training feature set was randomly divided in $ 5 $ equal parts. Each of the possible combinations of $ 4 $ parts was used to tune the model parameters, leaving the last part for fitting the hyperparameters by cross-validation. For each of the models, the grid search was performed for an initially large range for each hyperparameter. This range was increased until all optimal hyperparameters were inside the range, and not on the edge. Subsequently, the range was gradually decreased to home in on each optimal hyperparameter, until the performance plateaued. Each grid search lasted from a few seconds for simpler models like KNN to a few minutes for more complex models like MLP. The optimal hyperparameters selected during cross-validation are in Table \ref{tab3}.

%

\begin{table}[t]
	\caption{The hyperparameters of the ML models optimised with $ 5 $-fold cross-validation on the training dataset.}\label{tab3}
\begin{tabular}{|c|c|c|}
\hline
\textbf{Model} & \textbf{Hyperparameter} & \textbf{Value} \\ 
\hline
\hline
Decision Tree 	& Criterion & Gini impurity\\			
				& Maximum tree depth & $ 17 $\\
\hline
K-Nearest  & Train instances weighting & Inversely proportional to distance\\			
Neighbours			& Algorithm & Ball Tree\\
\hline
Support Vector  & Regularization parameter $ C $ & $ 316 $\\			
Machine		&	Kernel  & Radial basis function\\
		&	$ \gamma $ & $ 0.08 $\\
		&	Tolerance for stopping criterion & $ 0.0316 $\\
			\hline
Multi-Layer  & Hidden layer size & $ 18 $\\			
Perceptron		&	Activation function  & Hyperbolic tangent\\
	&		Algorithm & Quasi-Newton based optimiser\\
	&		Regularization parameter $ \alpha $ & $ 5\cdot10^{-5} $\\
	\hline
		\end{tabular}	
\end{table}

\subsection{Comparing with human-made heuristics}
\label{SUBSEC:Human}

The ML approaches were compared in terms of prediction accuracy and resulting CAD computing time against the two best known human-made heuristics.
\begin{description}
\item[Brown] This heuristic chooses a variable ordering according to the following criteria, starting with the first and breaking ties with successive ones:
\begin{enumerate}[(1)]
\item Eliminate a variable first if it has lower overall degree in the input.
\item Eliminate a variable first if it has lower (maximum) total degree of those terms in the input in which it occurs.
\item Eliminate a variable first if there is a smaller number of terms in the input  which contain the variable.
\end{enumerate}
It is named after Brown who documented it only in the notes of an ISSAC\footnote{\url{https://www.usna.edu/Users/cs/wcbrown/research/ISSAC04/handout.pdf}}.

\item[sotd] This heuristic constructs the full set of projection polynomials for each permitted ordering and selects the ordering whose corresponding set has the lowest \textbf{s}um \textbf{o}f \textbf{t}otal \textbf{d}egrees for each of the monomials in each of the polynomials. It was the reccommendation made after the study  \cite{DSS04}.
\end{description}  

Unlike the ML, these human-made heuristics can end up predicting several variable orderings (i.e. when they cannot discriminate).  In practice if this were to happen the heuristic would select one randomly (or perhaps lexicographically), however that final pick is not particularly meaningful for an evaluation.   To accommodate this, for each problem, the prediction accuracy of such a heuristic is judged to be the the percentage of its predicted variable orderings that are also target orderings.  The average of this percentage over all problems in the testing dataset represents the prediction accuracy.  Similarly, the computing time for such methods was assessed as the average computing time over all predicted orderings, and it is this that is summed up for all problems in the testing dataset.

\section{Results}
\label{SEC:Results}

We compare the four ML models on the percentage of problems where they selected the optimum ordering, and the total computation time (in seconds) for solving all the problems with their chosen orderings. We also compare the ML models with the two human-made heuristics (with the adaptations outlined in Section \ref{SUBSEC:Human}) and finally the outcome of a random choice between the 6 orderings. The results are presented in Table \ref{tab2}.  We might expect a random choice to be correct one sixth of the time for this data set (16.6\%).  The actual accuracy is a little higher because the dataset is not uniform, and for some problem instances there were multiple variable orderings with equally fast timings.

\begin{table}[b]
	\caption{The comparative performance of DT, KNN, MLP, SVM, and the Brown and sotd heuristics on the testing dataset.}\label{tab2}
	\begin{center}
		\begin{tabular}{|l|c|c|c|c|c|c|c|}
			\hline
			  &  DT & KNN & MLP & SVM & Brown & sotd & random\\
			\hline
			\textbf{Accuracy}  &  $ 62.6 \% $ & $ 63.3 \% $ & $ 61.6 \% $ & $ 58.8 \% $ & $ 51 \% $ & $ 49.5 \% $ & $ 22.7 \% $\\
			\textbf{Computation Time (s)} & $ 9,994 $ & $ 10,105 $ & $ 9,822 $ & $ 10,725 $ & $ 10,951 $ & $ 11,938 $ &  $ 30,235 $ \\			
			\hline
		\end{tabular}
	\end{center}
\end{table}

Moreover, we evaluate the distribution of the computation time for the ML methods and the heuristics. The differences between the computation time of each method and the minimum computation time, given as a percentage of the minimum time, are depicted in Figure 1.

\begin{figure}[t]
	\includegraphics[width=\textwidth]{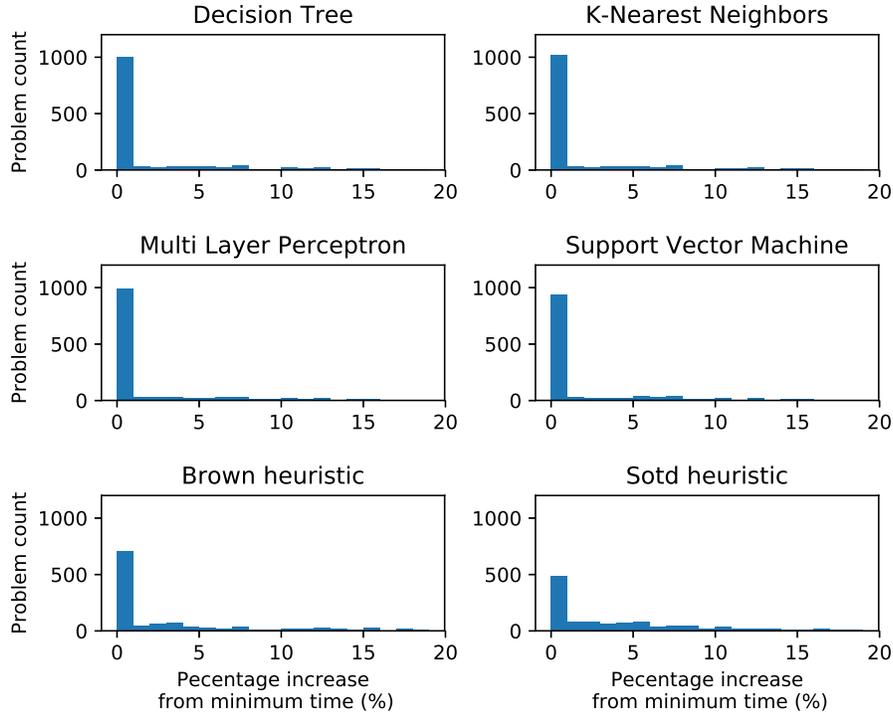}
	\caption{The histograms of the percentage increase in computation time relative to the minimum computation time for each method, calculated for a bin size of 1\%.} \label{fig1}
\end{figure}

\subsection{Range of possible outcomes}

We note that the minimum total computing time achieved by selecting an optimal variable ordering for every problem would be $8,623$s. Choosing at random would take $ 30,235 $s, almost 4 times as much. The maximum total computing time, determined by selecting the variable ordering with the longest computing time, is $ 64,534 $s. The choices with the quickest time among the methods considered were achieved by the Decision Tree model: $ 9,994 $s, which is $ 16\% $ more than the minimal possible. So there are clearly great time savings to be made by taking this choice into account.

\subsection{Human-made heuristics}

Of the two human-made heuristics, Brown performed the best, as it did in \cite{HEWDPB14}.  As was noted there this is surprising since the sotd heuristic has access to additional information (not just the input polynomials but also their projections). Unlike the ML models and the Brown heuristic, obtaining an ordering for a problem instance with sotd is not instantaneous.  Generating an ordering with sotd for all problems in the testing dataset took over $ 30 $min.

Brown could solve all problems in 10,951s, 27\% more than the minimum.  While sotd is only 0.7\% less accurate than Brown in identifying the best ordering, it is much slower at 11,938s or 38\% more than the minimum.  This shows that while Brown is not much better at identifying the best ordering, it does a much better job at discarding the worst!

\subsection{ML choices}

The results show that all ML approaches outperform the human-made heuristics in terms of both accuracy and timings.
Figure 1 shows that the human-made heuristics result in computing times that are often significantly larger than $ 1\% $ of the corresponding minimum time for each problem. The ML methods, on the other hand, all result in almost $ 1000 $ problems ($\sim 75\% $ of the testing dataset) within $ 1\% $ of the minimum time.

The key finding of the present paper is that there are significantly better models to use for the problem than SVM: each of the other three had 10\% higher accuracy and much lower timings.  
It is interesting to note that the MLP model leads to a lower accuracy than DT and KNN, but achieves the lowest overall computing time. This suggests that on this dataset, the MLP model is best at leaving out variable orderings that lead to long computing times, even if it has a slightly higher chance of missing the target variable ordering.  

\subsection{Conclusion}

For a casual user of computer algebra the key consideration is probably that their computation does not hang, and that it finishes reasonably fast on the widest variety of questions.  Hence of the tested models we may recommend the MLP for use going forward.  However, all the models show promise and we emphasise that this conclusion applies only for the present dataset and we need wider experimentation to see if this finding is replicated.

\section{Final Thoughts}
\label{SEC:End}

The experiment shows a clear advance on the prior state of the art:  both in terms of accuracy of predictions for CAD variable ordering; and in understanding which models are the best suited for this problem.  But we acknowledge that there is much more to do and emphasise that this is only the initial findings of the project.
The following extensions will be undertaken shortly:
\begin{itemize}
\item Extending the experiment to include problems with more variables from the dataset.  Unlike some CAD implementations the one used does not change algorithm when $n>3$; however, like all CAD implementation the time required will increase exponentially.

\item Using the trained classifiers to test on CAD problems from outside the dataset, in particular, those which are not fully existentially quantified (SAT problems).  In such cases the algorithms can change (terminate early) and it is not clear if models trained for the SAT case can be applied there.

\item Using the trained classifiers to test on CAD implementations other than the one in the \textsc{Maple} \texttt{RegularChains} Library.  For example, can the classifiers also usually pick variable orderings for \textsc{Qepcad-B}, \cite{Brown2003b}, or \textsc{Redlog} \cite{DS97a}.

\item Examining how best to derive additional features to use for the training, and the use of feature selections tools to find an optimal subset.

\item Test a dynamical selection of the variable ordering, where ML only picks the first variable for a problem, the polynomials are projected along that variable, and then the process repeats iteratively.

\end{itemize}
Finally, we note the wide variety of additional problems in computer algebra systems open to study with machine learning \cite{England2018}.

\subsection*{Acknowledgements}

The authors are supported by EPSRC Project EP/R019622/1: \emph{Embedding Machine Learning within Quantifier Elimination Procedures}.  

\subsubsection*{Research Data Statement} Data supporting the research in this paper is available at: \url{http://doi.org/10.5281/zenodo.2658626}.

\bibliographystyle{splncs04}
\bibliography{CAD}

\end{document}